\documentclass[twoside,12pt]{article}
\usepackage{graphicx}

\def\0{\over } \def\1{\mathbf } \def\2{{1\over2}} \def\4{{1\over4}}
\def\5{\bar }
\def\6{\partial }
\def\8#1{{\textstyle{#1}}}
\def\9#1{\underline{#1}}
\newcommand{\be}{\begin{equation}}
\newcommand{\ee}{\end{equation}}
\newcommand{\bea}{\begin{eqnarray}}
\newcommand{\eea}{\end{eqnarray}}

\begin{document}

\title{ \vspace{1cm} Hard loop effective theory of\\ the (anisotropic) quark gluon plasma}
\author{A.\ Rebhan
\\
\\
Institut f\"ur Theoretische Physik, Technische Universit\"at Wien,\\
        Wiedner Hauptstrasse 8-10, A-1040 Vienna, Austria}
\maketitle
\begin{abstract}
The generalization of the hard thermal loop effective theory to
anisotropic plasmas is described with a detailed discussion of
anisotropic dispersion laws and plasma instabilities.
The numerical results obtained in real-time lattice simulations
of the hard loop effective theory are reviewed, both for the
stationary anisotropic case and for a quark-gluon plasma
undergoing boost-invariant expansion.
\end{abstract}

\section{Introduction}

\subsection{\it Weakly vs.\ strongly coupled quark-gluon plasma}

The wealth of data harvested at the
Relativistic Heavy Ion Collider (RHIC) \cite{Tannenbaum:2006ch}
has led to a shift of paradigm in thinking about the quark-gluon plasma.
The strong collectivity that is being observed, in particular in
elliptic flow and jet quenching, is widely taken as pointing 
to a strongly coupled
plasma which is qualitatively and quantitatively different from a
parton plasma that can be described by perturbative quantum chromodynamics
(QCD); for an opposing point of view see however Ref.~\cite{Xu:2008dv}.

In fact, already before the mounting experimental evidence for a
strongly coupled quark gluon plasma (sQGP) at RHIC, perturbative QCD
at finite temperature was in some difficulty describing lattice
data on the thermodynamics of deconfined QCD. 
Lattice results for the thermodynamic pressure of QCD typically 
lead to a rather sudden rise of pressure and entropy to about
15-20 \% of the
Stefan-Boltzmann result for an ultrarelativistic gas of quarks and
gluons, at a few times the transition temperature $T_c$. A straightforward
first-order calculation in $\alpha_s$ in fact gives
just this ballpark of deviations from the interaction-free result.
Higher-order calculations require resummation of collective
effects such as Debye screening and have been carried through
to order $\alpha_s^3\ln\alpha_s$ \cite{Arnold:1994ps,Braaten:1996jr,Kajantie:2002wa}, but at face value they 
show hopelessly poor convergence
of perturbation theory at all temperatures of practical interest
(in fact up to ridiculously high temperatures $\sim 10^5 T_c$).

\begin{figure}[tb]
\begin{center}
\begin{minipage}[t]{8 cm}
{\includegraphics[width=\textwidth,bb=30 190 540 530,clip]{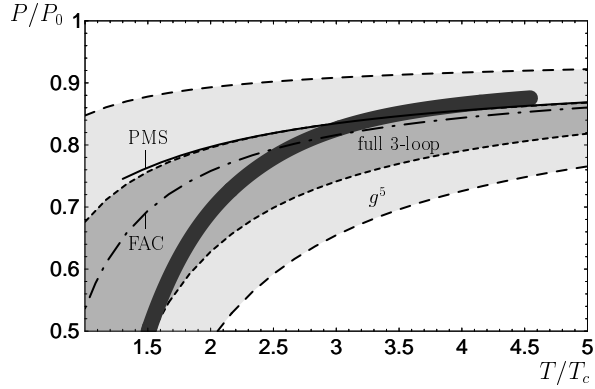}}
\end{minipage}
\begin{minipage}[t]{16.5 cm}
\caption{Comparison of the perturbative 3-loop result for the
pressure of pure-glue QCD with lattice results from \cite{Boyd:1996bx}
(thick dark line; thickness
representing roughly the statistical error). 
The light-gray band represents the result to order $g^5$
when the renormalization scale is varied between $\pi T$ and
$4\pi T$ and the medium-gray band shows the reduced renormalization scale
ambiguity when higher order terms are resummed
through the parameters of 
the effective field theory of dimensional reduction. The
lines marked ``PMS'' (for principle of minimal sensitivity)
and ``FAC'' (fastest apparent convergence) correspond to two
optimization prescriptions for the renormalization scale. 
\cite{Blaizot:2003iq} \label{fig3loop}
}
\end{minipage}
\end{center}
\end{figure}

\begin{figure}[tb]
\begin{center}

\begin{minipage}[t]{8 cm}
\centerline{
\raisebox{1.3mm}{\includegraphics[width=\textwidth]{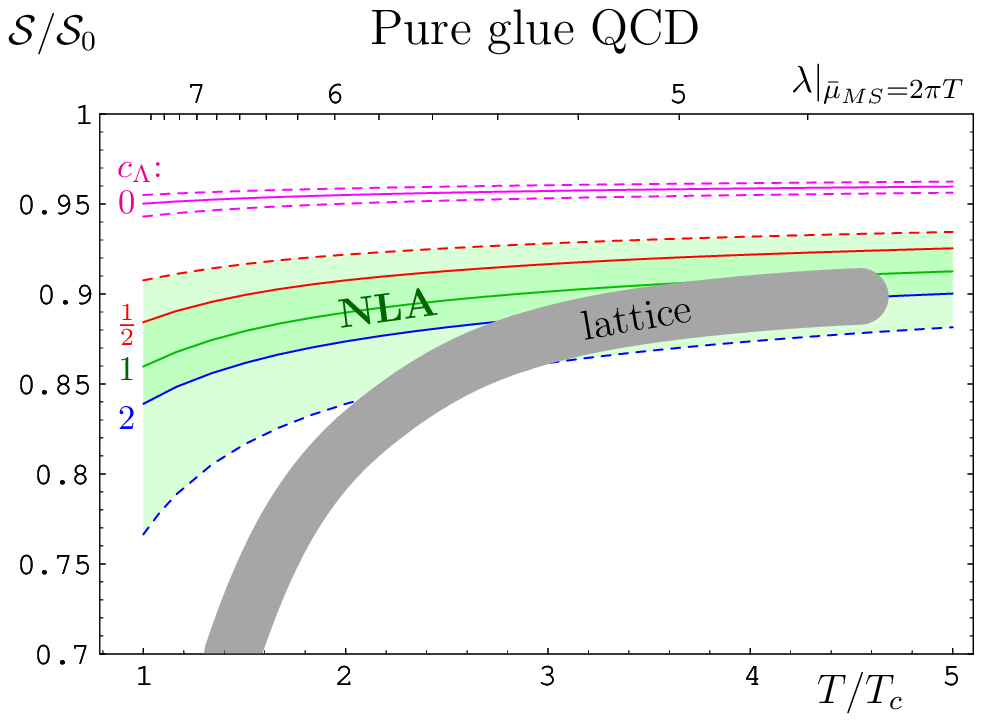}}
\includegraphics[width=0.9655\textwidth]{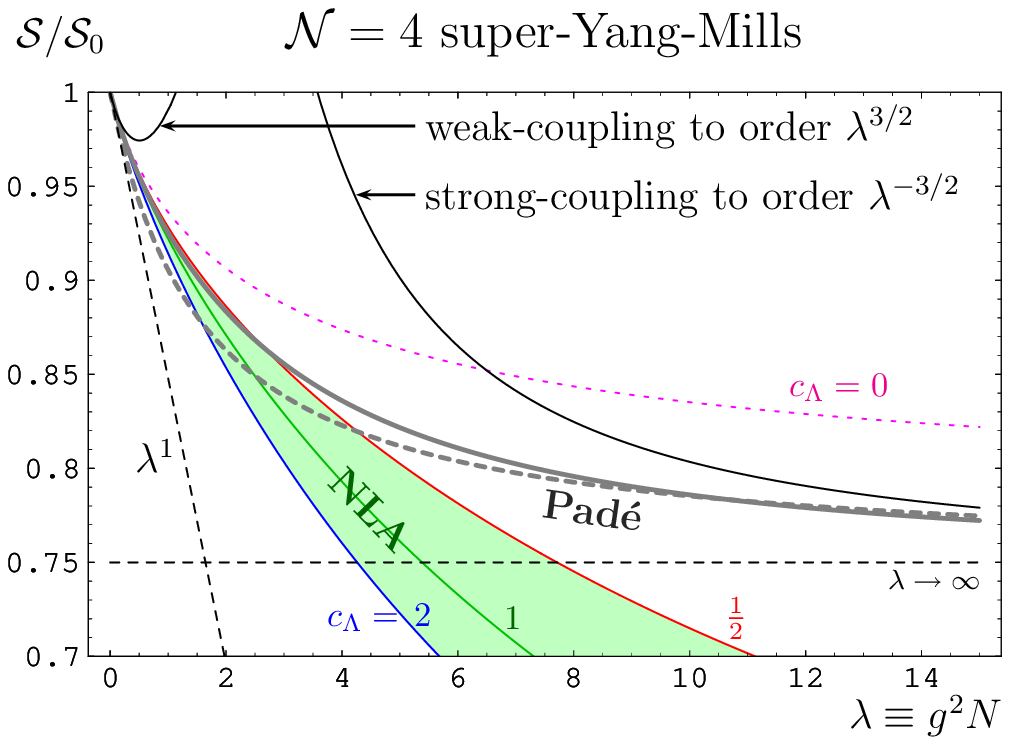}
}
\centerline{(a) \hfill (b)}
\end{minipage}
\begin{minipage}[t]{16.5 cm}
\caption{The left panel (a) compares the lattice result for
the entropy of pure-glue QCD of \cite{Boyd:1996bx}
with a next-to-leading HTL approximation (NLA)
where the quasi-particle entropy of HTL quasiparticles
is corrected by next-to-leading order corrections for
the asymptotic thermal masses in a simple model for a gap equation
(the parameter $c_\Lambda$ is introduced to
restrict this correction to hard modes above the scale
$\sqrt{2\pi T m_D c_\Lambda}$). The upper and lower
dashed lines correspond to renormalization scales $\bar\mu_{\rm
MS}=4\pi T$ and $\pi T$, respectively.
The right panel (b) shows the weak and strong coupling results for the entropy density
of $\mathcal N=4$ SYM theory together with the NLA results
obtained in analogy to QCD, but as a
function of $\lambda$, which here does not run.
The dashed and full heavy gray lines represent the Pad\'e
approximants $R_{[1,1]}$ and $R_{[4,4]}$ which interpolate
between weak and strong coupling results to leading and next-to-leading
orders, respectively \cite{Blaizot:2006tk}.
\label{fignla}
}
\end{minipage}
\end{center}
\end{figure}

However, it is now understood that the poor convergence properties of
thermal perturbation theory beyond first order perturbation theory
is at least to a large part
signalling the need for more complete resummations of
screening phenomena, since similar problems appear already in
the rather trivial case of scalar O($N\to\infty$) models, which
can be solved exactly and where the only effect is the
generation of a thermal mass \cite{Drummond:1997cw}.
Indeed, already a minimal resummation
of the Debye mass beyond strict perturbation theory together with
a simple optimizations of the (huge) renormalization
scale dependence \cite{Blaizot:2003iq}
(see Fig.~\ref{fig3loop}) gives a remarkably good
description of the (continuum-extrapolated) lattice results down
to about $2.5 T_c$. 
Fig.~\ref{fignla}a shows that a description
of the entropy of (pure glue) QCD in terms of hard-thermal-loop (HTL)
resummed quasiparticle propagators together with a standard
2-loop running coupling $\alpha_s$ (with renormalization scale
varied about the Matsubara scale $2\pi T$) also gives a good description
of the thermodynamics above about $3T_c$ \cite{Blaizot:2000fc} (see also
Ref.~\cite{Andersen:2004fp} and references therein).\footnote{HTL quasiparticle
models have also been used as phenomenological models down to
the phase transition by adding fitting parameters in the
running coupling \cite{Peshier:1996ty,Rebhan:2003wn,Schulze:2007ac}, 
sometimes extended by
incorporating quasiparticle damping in a form motivated by
HTL perturbation theory \cite{Peshier:2004bv,Cassing:2007nb}.}

Meanwhile, new techniques have become available that allow the analytical
treatment of strongly coupled gauge theories.  In particular,
$\mathcal N=4$ supersymmetric Yang-Mills theories (SYM) in the limit
of large number of colors $N$ and strong 't Hooft coupling
$\lambda\equiv g^2 N=4\pi\alpha_s N$ is now often taken as a model for hot QCD.
On the basis of 
the AdS/CFT conjecture \cite{Maldacena:1997re,Aharony:1999ti} one
can make predictions for the otherwise inaccessible strong-coupling
regime, notably for real-time quantities such as
the specific shear viscosity $\eta/\mathcal S$. While $\eta/\mathcal S \sim
(\lambda^2\log\lambda)^{-1}\gg 1$ at weak coupling
\cite{Arnold:2003zc,Huot:2006ys}, the AdS/CFT correspondence
gives $\eta/\mathcal S=1/4\pi+O(\lambda^{-3/2})$ at large 't Hooft
coupling \cite{Policastro:2001yc,Buchel:2008sh}, and such (extremely) 
low values
seem to be indeed required in viscous hydrodynamics models of heavy-ion
collisions at RHIC \cite{Romatschke:2007mq} 

For $\mathcal N=4$ SYM at finite temperature, no lattice results for
the thermodynamic potential are available that would allow one to
test either weak coupling results or the (conjectured) strong coupling
results at finite coupling (which does not run because of conformal
invariance). However, successive Pad\'e approximants that interpolate
between the known weak and strong coupling results give smooth and
seemingly robust extrapolations, as shown in Fig.~\ref{fignla}b
\cite{Blaizot:2006tk}. A comparison of 
the QCD results for the entropy with those for $\mathcal N=4$ SYM
suggests that the strong-coupling expansion is no longer working
well when the deviations from the Stefan-Boltzmann result ($\mathcal S_0$)
are less than some 15\%, as is the case for QCD at temperatures
above $3 T_c$. Such temperatures are expected to be reached in
the heavy-ion experiments at the LHC, which may eventually get
to probe specific perturbative features of the quark-gluon plasma.
But already in the case of RHIC physics, it is clearly
mandatory to improve our understanding of both the weak and the
strong coupling asymptotics of the quark-gluon plasma, as the
truth will most probably be somewhere in between, perhaps
quite far from either. 

In some respect, a weakly coupled quark-gluon plasma (wQGP) can be
more difficult to describe than a strongly coupled one,
in which strong coupling dynamics wipes out any structures
from quasiparticle dynamics. In a wQGP, a small coupling
leads to a hierarchy of scales, which need to be disentangled
by assuming $g\ll 1$. Of course, for quantitative predictions,
extrapolations to $g\sim 1$ have to be considered, but this
is no more or less problematic than the equally bold extrapolations
theoretician need to do by starting out from infinite ('t Hooft)
coupling.

\subsection{\it Scales of wQGP}

In thermal equilibrium, the primary scale is the temperature $T$,
which determines the mean energy of (``hard'') particles in the plasma.
The ``soft'' scale $gT$ with $g\ll1$ is the scale of thermal
masses that determine the plasma frequency, below which no
propagating modes exist, the Debye screening mass, and the scale
of Landau damping. In Feynman diagrams, these effects come from
one-loop contributions with the highest power of the loop momentum,
cut off by the thermal distribution function, and are therefore
called hard thermal loops (HTL). To determine the physics at
softer scales, one typically needs to consider the effective
HTL theory and its resummed propagators and vertices.

At the scale $g^2T$ one encounters in fact a barrier for perturbation
theory when the magnetostatic sector is involved, since the latter
is characterized by a completely nonperturbative dimensionally
reduced Yang-Mills theory, which by itself can be (comparatively
easily) handled by lattice methods (the difficult part being
the perturbative matching to the four-dimensional theory).
Quantities such as the color relaxation or gluon damping
rate \cite{Pisarski:1993rf,Flechsig:1995sk} and also the Debye screening
length at next-to-leading order \cite{Rebhan:1993az}
involves logarithms of the coupling, whose coefficient can
be calculated perturbatively, but which are nonperturbative
beyond the leading log.

The scale $g^4T$ is the one characteristic of large-angle scattering
and also of the inverse shear viscosity $T^4/\eta$, whose
calculation at weak coupling \cite{Arnold:2003zc,Huot:2006ys}
requires resummations beyond,
but building upon, hard thermal loops.

When it comes to nonequilibrium physics, it turns out that
the parametrically most important phenomena are
plasma instabilities 
\cite{Mrowczynski:1988dz+Mrowczynski:1993qm,Pokrovsky:1988bm}
that appear already in the collisionless
limit, on the level of the HTL masses (and only those of
gauge bosons \cite{Schenke:2006fz}). As pointed out
by Arnold et al.\ \cite{Arnold:2003rq}, this
leads for instance to the necessity of a complete revision
(still to be worked out) for systematic perturbative scenarios
of thermalization \cite{Baier:2000sb}. Intriguingly,
plasma instabilities may also be responsible for anomalous
contributions \cite{Asakawa:2006tc} to the inverse shear viscosity whose
standard weak coupling result appears much too small (which
is one of the key arguments in favor of sQGP).

\section{Hard thermal and hard anisotropic loops}

\subsection{\it Hard (thermal) loop effective theory}

The effective HTL theory of QCD \cite{Frenkel:1990br,Braaten:1990mz}
is given by the collection of all
one-loop diagrams which are proportional to $T^2$, assuming that
all external momenta are soft and therefore negligible compared to $T$.
For external momenta $\sim gT$, the HTL self energy and vertex diagrams
are parametrically of the same order as the corresponding tree-level
quantities and therefore need to be resummed completely. A remarkably
compact form of the HTL effective action can be written down formally
\cite{Braaten:1992gm}, which for gluons reads
\be\label{BPeffact}
{\cal L}_{\rm HTL}(x) = 
{m_D^2\0 2}
\Big\langle 
F_{\mu \nu}^a (x)
\bigg({ v^\nu  p^\rho \over ( v \cdot D)^2} \bigg)_{ab} \;
F_\rho^{b \,\mu} (x) 
\Big\rangle_{{\bf v}} \;.
\ee
Here $m_D\sim gT$ is the Debye mass in the HTL approximation,
$\langle\cdots\rangle_{{\bf v}}$ a normalized
average over the directions $\vec v$ in $v^\mu=(1,\vec v)$ with $\vec v^2=1$, and
$D_\mu$ the covariant derivative in the adjoint representation.
In the non-Abelian case, this leads to an effective action
which is nonlocal and nonpolynomial,
containing an infinity of vertex functions. The reason for
nonlocality is that, in contrast to other examples of an effective
field theory, one is integrating out stable real particles rather
than virtual ones. Albeit elegant in form, the effective action
(\ref{BPeffact}) is not well-defined as it stands, because it still
requires boundary conditions to be taken into account after
the extraction of self energy and vertex diagrams.

Alternatively, the HTL effective theory can be derived from
the effective field equation
\be\label{DFj}
D_\mu F^{\mu\nu}_a=j^\nu_a[A]=
g \int{d^3p\0(2\pi)^3} \frac{p^\mu}{2p^0} \delta f_a(\1p,\1x,t)\,,
\ee
where $\delta f_a$ is a linearized perturbation of
the color-neutral background (thermal)
distribution of color-carrying hard particles
$v\cdot\partial\, f_0(\1p,\1x,t)=0$, $v^\mu=p^\mu/p^0$,
satisfying
gauge covariant Boltzmann-Vlasov equations \cite{Blaizot:2001nr}
\be\label{vDf}
v\cdot D\, \delta f_a(\1p,\1x,t)=g v_\mu F^{\mu\nu}_a \6^{(p)}_\nu f_0(\1p,\1x,t)=-g(\1E_a+\1v\times\1B_a)\cdot\nabla_{\1p}f_0\,.
\ee
HTL vertex functions are then defined as the functional derivatives
of the induced current $j[A]$ with respect to $A_\mu^a$. In particular,
the HTL gauge boson self energy is given by the linear terms in $A_\mu^a$,
which in Fourier space and imposing retarded boundary conditions reads
\be\label{Pimunu}
j^{\mu}(k)|_{\rm lin.}=g^2 \int \frac{d^3 p}{(2\pi)^3} v^{\mu} %
\partial^{\beta}_{(p)} f_0({\bf p}) 
\left( g_{\gamma \beta} - %
\frac{v_{\gamma} k_{\beta}}{k\cdot v + i \epsilon}\right) A^{\gamma}(k) =
\Pi^{\mu\nu}(k)A_\nu(k)\,.
\ee

Written as gauge-covariant Boltzmann-Vlasov equations, the HTL effective
theory allows us to immediately generalize to nonthermal
cases. When the background distribution function $f_0$ is stationary
and homogeneous, i.e.\ only dependent on momentum, one can
in fact still write down an effective action similar to (\ref{BPeffact})
\cite{Mrowczynski:2004kv}. The main difference to the
thermal situation is that $\nabla_{\1p}f_0$ need no longer be
proportional to $\mathbf p$ and thus $\mathbf v$. Whereas in
the thermal (or just isotropic) case, the magnetic term in the Vlasov equation (\ref{vDf}) drops out, in the anisotropic case magnetic interactions
lead to much more complicated and rich dynamics.

\subsection{\it Isotropic dispersion laws}

\begin{figure}[tb]
\begin{center}
\begin{minipage}[t]{8 cm}
{\includegraphics[width=\textwidth,bb=34 124 516 450,clip]{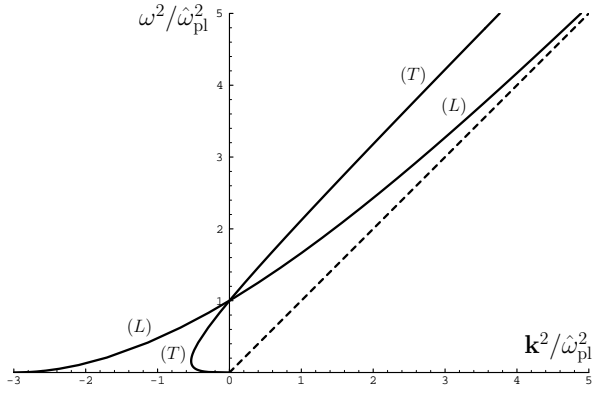}}
\end{minipage}
\begin{minipage}[t]{16.5 cm}
\caption{Location of poles in $\Delta_T$ and $\Delta_L$. 
The right part with
$\mathbf k^2\ge0$ corresponds to propagating normal modes,
the left part to (dynamical) screening. \label{figg}
}
\end{minipage}
\end{center}
\end{figure}

In the isotropic case, the polarization tensor $\Pi^{\mu\nu}$
in (\ref{Pimunu}), which is transverse with respect to the 4-momentum $k$,
contains two structure functions corresponding to the two
symmetric and transverse tensors
$A_{\mu\nu}=g_{\mu\nu}-{k_\mu k_\nu \0 k^2} - B_{\mu\nu}$, 
$B_{\mu\nu}={\tilde u_\mu \tilde u_\nu \0 \tilde u^2}$ with
$\tilde u_\mu=(g_{\mu\nu}-{k_\mu k_\nu \0 k^2})u^\nu$
and plasma rest-frame velocity $u^\mu=\delta^\mu_0$,
\bea
&&\Pi_A\equiv
\Pi_T={1\02}A_{\mu\nu}\Pi^{\mu\nu}={1\02}\left( \Pi^\mu{}_\mu-\Pi_B \right),
\qquad\Pi_B\equiv \Pi_L=-{k^2\0\mathbf k^2}\Pi_{00}\\
&&\Pi^\mu{}_\mu=m_D^2,\qquad \Pi_{00}=m_D^2
\left(1-{k^0\02|{\bf k}|}\ln{k^0+|{\bf k}|\0k^0-|{\bf k}|}\right).
\eea
As a consequence, 
the gauge boson propagator, which in Landau gauge reads
\be
-G_{\mu\nu}=\Delta_T A_{\mu\nu} + \Delta_L B_{\mu\nu},\qquad
\Delta_T=[k^2-\Pi_T]^{-1},\quad\Delta_L=[k^2-\Pi_L]^{-1},
\ee
has two branches of poles, corresponding to two different dispersion
laws, $\Delta_T^{-1}=0$ for spatially transverse polarizations,
and $\Delta_L^{-1}=0$ for polarizations along $\tilde u$.

The two branches of poles are shown in Fig.~\ref{figg},
plotted in $\omega^2$ over $\mathbf k^2$ in order to show
poles corresponding to propagating modes ($\mathbf k^2>0$)
and poles corresponding to screening ($\mathbf k^2<0$), which
are analytically connected, on the
same plot. Propagating modes are seen to exist for $\omega^2\ge
\hat\omega_{\rm pl}^2=m_D^2/3$, with the transverse mode
approaching the mass hyperboloid $k^2=m_\infty^2=m_D^2/2$
asymptotically. The poles of branch $L$ approach the light-cone exponentially,
and in doing so, their residues vanish also exponentially,
signifying a purely collective mode. For frequencies below
the plasma frequency, $\omega<\hat\omega_{\rm pl}$, there is
screening of both electric and magnetic fields, with screening
lengths depending on frequency. In the static limit, mode $L$
has inverse screening length $m_D$, the so-called Debye mass.
This corresponds to the screening of (chromo-)electric charges
in the medium. By contrast, mode $T$ has infinite screening length
in the limit of vanishing frequency, corresponding to the
absence of (chromo-)magnetostatic screening in the
HTL approximation.

\subsection{\it Anisotropic dispersion laws and plasma instabilities \label{secansio}}

In the anisotropic case, the fact that 
$\Pi^{\mu\nu}$ is symmetric and $\Pi^{0\nu}$ fixed by transversality $k_\mu \Pi^{\mu\nu}=0$ would lead to 6 structure functions in general.
Assuming that there is just one direction of momentum space anisotropy,
$\mathbf n=(0,0,1)$ (i.e., axisymmetry around the $z$-axis), one
can define
4 symmetric tensors for $\Pi^{ij}$, corresponding to
4 independent structure functions.

Defining spatial tensors
\be
A^{ij}=\delta^{ij}-k^{i}k^{j}/k^2,\quad B^{ij}=k^{i}k^{j}/k^2,
\quad C^{ij}=\tilde{n}^{i} \tilde{n}^{j} / \tilde{n}^2,\quad
D^{ij}=k^{i}\tilde{n}^{j}+k^{j}\tilde{n}^{i},
\ee
with $\tilde{n}^{i}=A^{ij} n^{j}$
and decomposing the spatial part of $\Pi^{\mu\nu}$ according to
\be
\Pi^{ij}=\alpha A^{i j}+\beta B^{ij} + \gamma C^{ij} + \delta D^{ij}
\ee
one finds that the gluon propagator in temporal axial gauge
reads \cite{Romatschke:2003ms+Romatschke:2004jh}
\bea
&&{\bf \Delta}(k) =  \Delta_T {\bf A} + (\mathbf k^2 - \omega^2 + \alpha + \gamma)\Delta_{\mathcal L} {\bf B}
              + [(\beta-\omega^2)\Delta_{\mathcal L} - \Delta_T] {\bf C} - \delta \Delta_{\mathcal L} {\bf D}
\\
&&
\Delta_T(k) = [\mathbf k^2 - \omega^2 + \alpha]^{-1},
\qquad
\Delta_{\mathcal L}(k) = [(\mathbf k^2 - \omega^2 + \alpha + \gamma)(\beta-\omega^2)-\mathbf k^2 \tilde n^2 \delta^2]^{-1}.
\eea
This propagator contains one branch of poles from $\Delta_T^{-1}=0$,
and in general two branches from $\Delta_{\mathcal L}^{-1}=0$, except
when $\mathbf k \parallel \mathbf n$ and thus $\tilde n=0$, in which
case there are only two branches in total.

\begin{figure}[tb]
\begin{center}
\begin{minipage}[t]{8 cm}
{\includegraphics[width=\textwidth]{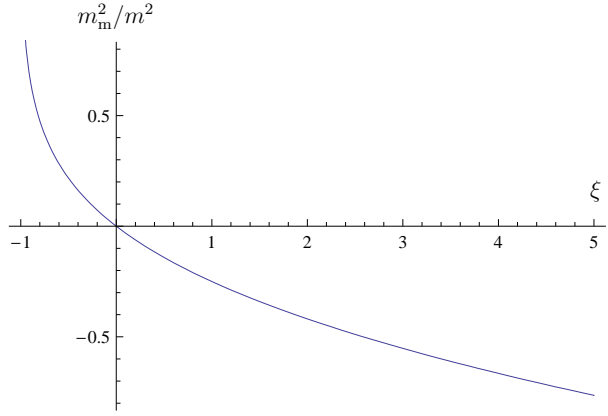}}
\end{minipage}
\begin{minipage}[t]{16.5 cm}
\caption{The magnetostatic screening mass squared as a function
of the anisotropy parameter $\xi$ as given in (\ref{mm2}).\label{figmm}
}
\end{minipage}
\end{center}
\end{figure}

A special important case for an axisymmetric distribution function
is given by
\be\label{fxi}
f({\bf p}) = f_{\rm iso}\left({\bf p}^2+\xi({\bf p}\cdot{\bf n})^2\right)
\ee
with some isotropic (for instance thermal) function $f_{\rm iso}$.
The anisotropy parameter $\xi$ is allowed to take values from $-1$ to
$\infty$, with $-1<\xi<0$ corresponding to prolate (cigar-shaped)
momentum anisotropy, and $0<\xi<\infty$ corresponding to oblate (squashed)
distributions. The polarization tensor in the
(anisotropic) hard loop (HL) approximation (\ref{Pimunu}) can then
be evaluated in closed form \cite{Romatschke:2003ms+Romatschke:2004jh}.
Changing variables ${\bf p}^2+\xi({\bf p}\cdot{\bf n})^2=\bar \mathbf p^2$
gives
\bea
\Pi^{i j}(k) &=& m^2 \int \frac{d \Omega}{4 \pi} v^{i}%
\frac{v^{l}+\xi({\bf v}.{\bf n}) n^{l}}{%
(1+\xi({\bf v}.{\bf n})^2)^2}
\left( \delta^{j l}+\frac{v^{j} k^{l}}{k\cdot v + i \epsilon}\right)\\
&&m^2 \equiv -{g^2\over 2\pi^2} \int_0^\infty d \bar p \,  
  \bar p^2 {d f_{\rm iso}(\bar p^2) \over d\bar p}.\nonumber
\eea

As a simple special case, let us consider structure function $\alpha(k)$
for the case that $\mathbf k \parallel \mathbf n$ and in the static limit 
where $\alpha(\omega\!=\!0)=
\2\Pi^{ii}(\omega\!=\!0)=m_{\rm m}^2$ has the interpretation of
magnetostatic screening mass squared.
One easily finds
\be\label{mm2}
{m_{\rm m}^2\0m^2}\Bigg|_{\mathbf k \parallel \mathbf n} = \left\{
{ {1\04}\left[(1-\xi)({-\xi})^{-1/2}{\rm atanh}({-\xi})^{1/2}-1\right]  \mbox{\quad for $\xi<0$} \atop
 {1\04}\left[(1-\xi)\xi^{-1/2}{\arctan \xi^{1/2}}-1\right]  \mbox{\qquad\qquad for $\xi>0$}
}
\right.
\ee
which is plotted in Fig.~\ref{figmm}.

For $\xi=0$, $m_{\rm m}^2$ vanishes, reproducing the result that
there is no HTL magnetostatic screening mass.
However, for $\xi<0$ (prolate momentum anisotropy) we obtain
a nonvanishing HL magnetostatic screening mass, and for $\xi>0$
this mass turns out to be imaginary, signalling an instability.

\begin{figure}[tb]
\begin{center}
\begin{minipage}[t]{7 cm}
{\includegraphics[width=\textwidth,clip]{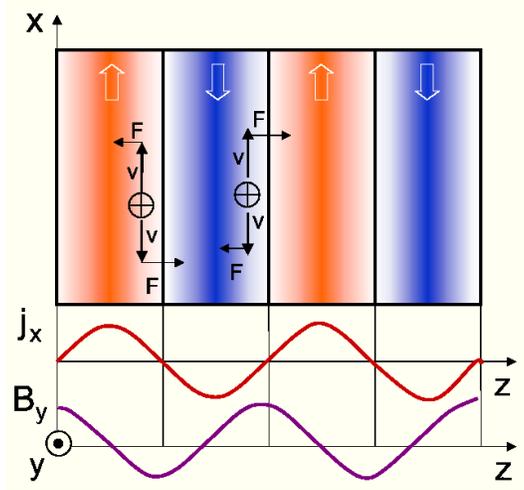}}
\end{minipage}
\begin{minipage}[t]{16.5 cm}
\caption{Illustration of the mechanism of filamentation (Weibel) instabilities
(adapted from Ref.\ \cite{Mrowczynski:2005ki}). Charged plasma particles moving transversely to the wave vector of
a seed magnetic field tend to separate in counterstreaming currents which
in turn amplify the magnetic field.  \label{figM1}
}
\end{minipage}
\end{center}
\end{figure}

This magnetic instability can be identified with the so-called
Weibel instability \cite{Weibel:1959} and is illustrated in Fig.~\ref{figM1}.
In an extremely oblate momentum distribution, where all hard particles
move in planes orthogonal to the $z$-axis, a small seed magnetic field
with wave vector in the $z$-direction tends to sort streams of charged
particles in a way which leads to an induced current that reinforces
the initial field. The induced current and the magnetic field
grow exponentially, until the magnetic field
is large enough to bend the trajectories of the hard particles
significantly, which in an Abelian plasma provides a mechanism
for fast isotropization. Of course, in the HL approximation, we
can only study the onset of such instabilities, because
the backreaction on the background distribution $f_0$ is neglected.

In fact, there are also instabilities when $\xi<0$, but to see those
we have to consider wave vectors pointing away from the $z$-axis.
For small $\xi$, one finds that in the static limit
there are the following
poles for real wave vector $\mathbf k$ with $k_z/|\mathbf k|
\equiv \cos \theta$,
\bea
\Delta_T^{-1}=0:&&\qquad \mathbf k^2/m^2 = \frac{1}{3}\,\xi\, \cos^2\theta 
+ O(\xi^2), \\
\Delta_{\mathcal L}^{-1}=0:&&\qquad \mathbf k^2/m^2 = 
\frac{1}{3}\,\xi\, \cos 2\theta + O(\xi^2).
\eea
Thus, when $\xi<0$, there are space-like poles in $\Delta_{\mathcal L}$ when 
$ \pi/4 < \theta < 3\pi/4$, corresponding to electric (Buneman) instabilities
\cite{Arnold:2003rq}.

\begin{figure}[tb]
\begin{center}
\begin{minipage}[t]{9.5 cm}
{\includegraphics[width=\textwidth]{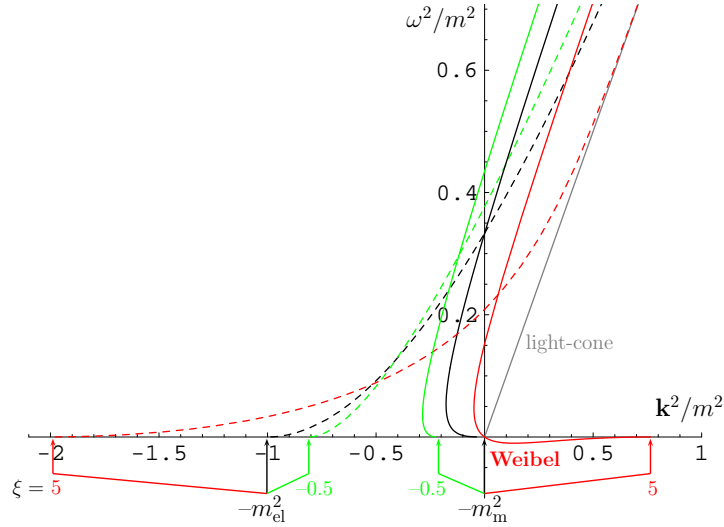}}
\end{minipage}
\begin{minipage}[t]{16.5 cm}
\caption{Anisotropic dispersion laws for $\mathbf k \parallel \mathbf n$
and three values of $\xi$: 0 (isotropic), $-0.5$ (prolate), and 5 (oblate).
Full and dashed lines represent poles of $\Delta_T$ and $\Delta_{\mathcal L}$,
respectively. The magnetic Weibel instability corresponds to the
poles for negative $\omega^2$ and positive $\mathbf k^2$ that appear
for positive $\xi$. 
\label{figdispan}
}
\end{minipage}
\end{center}
\end{figure}

Returning to the case of oblate distribution functions, $\xi>0$,
and restricting again 
to $\mathbf k \parallel \mathbf n$ (i.e., $|\cos\theta|=1$),
which gives the largest tachyonic mass, let us now consider
the dependence on frequency in order to determine the
full dispersion laws. With $\eta=\omega/|\mathbf k|$ one finds 
\cite{Romatschke:2003ms+Romatschke:2004jh}
\bea
\alpha(\eta)&=&\frac{m^2}{4 \sqrt{\xi} (1+\xi \eta^2)^2} %
\left[\left(1+\eta^2+\xi (-1+(6+\xi) \eta^2-(1-\xi) \eta^4)\right) %
\arctan \sqrt{\xi}\right.\nonumber \\
&& \hspace{4cm} \left.+ \sqrt{\xi}\, (\eta^2-1) 
\left(1+\xi \eta^2-(1+\xi) \eta \ln{\frac{%
\eta+1+i \epsilon}{\eta-1+i \epsilon}}\right)\right],\\
\beta(\eta)&=&- \frac{\eta^2 m^2}{2 \sqrt{\xi} (1+\xi \eta^2)^2} %
\left[(1+\xi)(1-\xi \eta^2)\arctan{\sqrt{\xi}} \right. \nonumber \\
&& \hspace{4cm} \left. + \sqrt{\xi} \left(%
(1+\xi \eta^2)-(1+\xi) \eta \ln{\frac{\eta+1+i \epsilon}{\eta-1+i \epsilon}}
\right)\right].
\eea
The resulting poles in $\Delta_T$ and $\Delta_{\mathcal L}$ are
plotted in Fig.~\ref{figdispan} by full and dashed lines, respectively,
for the three cases $\xi=0$ (isotropic), $\xi=-0.5$ (prolate), and
$\xi=5$ (oblate).

\begin{figure}[tb]
\begin{center}
\begin{minipage}[t]{8 cm}
{\includegraphics[width=\textwidth]{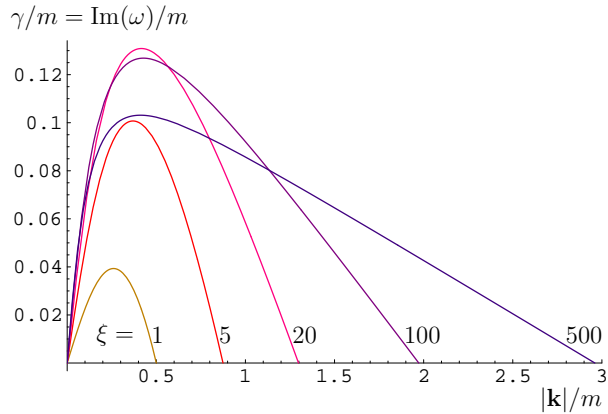}}
\end{minipage}
\begin{minipage}[t]{16.5 cm}
\caption{The growth rate $\gamma$ of the unstable modes
in a Weibel instability as function of wave vector and for several
values of anisotropy parameter $\xi$. \label{figgamma}
}
\end{minipage}
\end{center}
\end{figure}

In the prolate case, $\xi=-0.5$, we see that there is magnetic
screening down to and including the static limit, whereas the
electrostatic screening is somewhat diminished (for fixed $m$).
In the oblate case, $\xi=5$, the latter is instead enlarged 
but there is now a pole at zero frequency and positive $\mathbf k^2$,
which is analytically connected with the dynamical magnetic screening
poles at $\mathbf k^2<0$ by a line of poles with real $\mathbf k$
and negative $\omega^2$. The corresponding imaginary values
give the momentum-dependent growth rate $\gamma(k)$ of the
unstable magnetic modes. These growth rates are shown in more detail
in Fig.~\ref{figgamma} for various values of $\xi$.

One can show that in the limit of large $\xi$, the unstable modes
are characterized by
$k_{\rm max}/m\sim {\xi^{1/4}}$, but 
$k/m|_{\gamma=\gamma_{\rm max}}\sim 1$.
Compared to the asymptotic gluon mass $m_\infty$
one has 
$k_{\rm max}/m_\infty\sim\sqrt\xi$ 
and $\gamma_{\rm max}/m_\infty \to 1/\sqrt2$.

\section{Discretized hard-loop effective theory}

In the Abelian case, the above dispersion laws give already a
complete overview of the stable and unstable modes
to leading order in the HL approximation. In particular,
the unstable modes grow exponentially with rate $\gamma(k)$
until finally the limit of the HL approximation is reached
and backreaction on the hard particle distribution $f_0$
has to be considered.
However, with non-Abelian
fields, the HL effective theory involves infinitely many vertex
functions which become important when unstable modes have grown
to amplitudes so large that the two terms in the covariant derivative
$\6_\mu-igA_\mu$ are comparable, i.e., when $gA\sim m$.
This is still within the HL approximation, since hard particles
have momentum $m/g$ and so backreaction on them becomes important
only when $gA\sim m/g \gg m$. It is therefore an important question
whether non-Abelian plasma instabilities are able to grow
beyond the intrinsically non-Abelian regime $gA\sim m$,
and this question is one that can be answered entirely within
the HL approximation.

In order to cope with the nonlocal structure of the HL effective
theory, it is useful to introduce an auxiliary field formulation as
developed in the HTL case in \cite{Nair:1993rx,Blaizot:1994am}.
Factorizing
\be
\delta f^a(x;p)=-g W^a_\mu(t,\mathbf x;\mathbf v)
\6_{(p)}^\mu f_0(\mathbf p)
\ee
one has
\be
[v\cdot D(A)]W_\mu(x;\mathbf v)
= F_{\mu\gamma}(A) v^\gamma 
\ee
and
\be
D_\rho(A)F^{\rho\mu}=j^\mu(x) = -g^2
\int {d^3p\over(2\pi)^3} 
{1\over2|\mathbf p|} \,p^\mu\, {\partial f(\mathbf p) \over \partial p^\nu}
W^\nu(x;\mathbf v).
\ee
In the latter equation, only one linear combination of the four fields
$W^\nu(x;\mathbf v)$ participates in the dynamical evolution.
This extra field is a field on both, configuration space and velocity
space, but in terms of this field together with the gauge potential,
the dynamical equations are local and polynomial.
These equations can therefore be solved by real-time lattice simulations
upon discretizing both spacetime and velocity space. The induced current
can then be written simply as 
\be
j^\mu(x) = {1\0\mathcal N}\sum\limits_{\mathbf v} v^\mu 
\mathcal W_{\mathbf v}(x)
\ee
by introducing $\mathcal N$ vectors $\mathbf v$ covering the unit
sphere. In the HTL case, such a discretization has been studied
first in \cite{Rajantie:1999mp}, whereas \cite{Bodeker:1999gx}
performed lattice simulations involving truncated series of
spherical harmonics for that purpose.

\begin{figure}[tb]
\begin{center}
\begin{minipage}[t]{8 cm}
{\includegraphics[width=\textwidth]{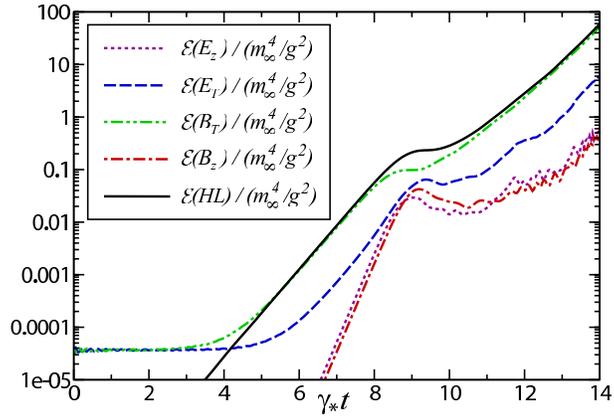}}
\end{minipage}
\begin{minipage}[t]{16.5 cm}
\caption{Lattice simulation of the non-Abelian
Weibel instability with small initial seed fields
that are constant in the transverse direction \cite{Rebhan:2004ur}.
Plotted are the average energy densities $\mathcal E$
in transverse/longitudinal chromomagnetic/electric
fields and the total energy density transferred from
hard particles to soft modes, $\mathcal E({\rm HL})$.
\label{figdhl}
}
\end{minipage}
\end{center}
\end{figure}

The first lattice study of non-Abelian plasma instabilities using
discretized hard loops was performed in \cite{Rebhan:2004ur}
for modes with $\mathbf k \parallel \mathbf n$.
The restriction to modes which are constant with respect to
spatial directions perpendicular to $\mathbf n$ leads to the great
simplification of a dimensional reduction to 1+1 spacetime
dimensions that need to be discretized in addition to the
2-dimensional (compact) velocity space. The results obtained
showed complicated nonlinear dynamics when
Weibel instabilities enter the intrinsically non-Abelian regime,
but except for a brief period of stagnation, it appeared
that non-Abelian plasma instabilities continue to grow exponentially
also in the late phase of the HL evolution, suggesting that
they are essentially as effective for fast isotropization as their Abelian
counterparts in conventional plasma physics. Closer inspection
reveals that this continued growth is brought about by an
effective Abelianization of the non-Abelian modes over
extended (but finite) spatial domains.

\begin{figure}[tb]
\begin{center}
\begin{minipage}[t]{8 cm}
\centerline{\includegraphics[width=\textwidth]{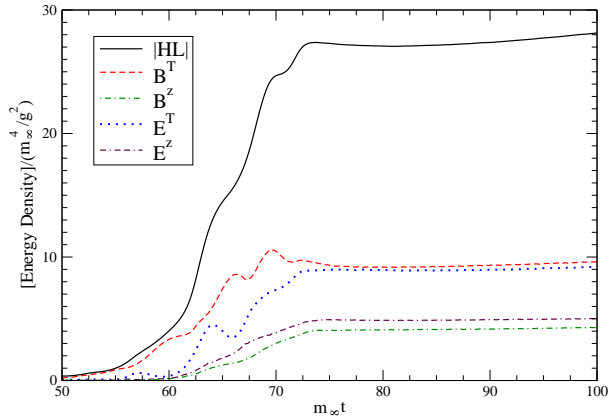}}
\end{minipage}
\begin{minipage}[t]{16.5 cm}
\caption{Lattice simulation of the non-Abelian Weibel instability as in
Fig.\ \ref{figdhl} but now with generic, fully 3-dimensional 
initial seed fields and
in linear scale, showing a saturation of exponential growth
in the strong-field regime \cite{Rebhan:2005re}.
\label{fig3d}
}
\end{minipage}
\end{center}
\end{figure}

\begin{figure}[tb]
\begin{center}
\begin{minipage}[t]{8 cm}
\centerline{\includegraphics[width=\textwidth,clip]{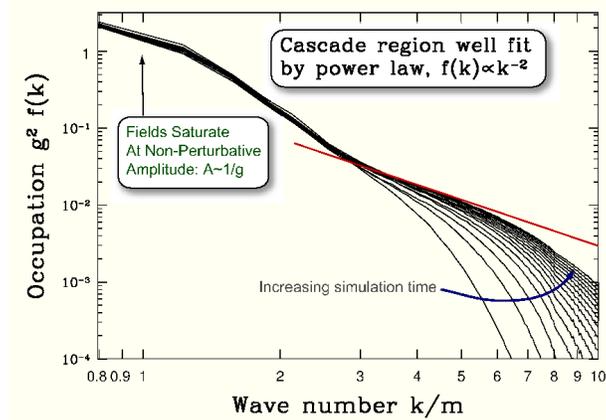}}
\end{minipage}
\begin{minipage}[t]{16.5 cm}
\caption{Spectrum of field modes showing a saturation at low wave numbers
and a cascading of energy to larger wave numbers with increasing
simulation time (adapted from Ref.~\cite{Arnold:2005ef}).
\label{figcasc}
}
\end{minipage}
\end{center}
\end{figure}

However, subsequent fully 3+1-dimensional lattice simulations
\cite{Arnold:2005vb,Rebhan:2005re}
showed that this phenomenon is only true for the modes
with exactly $\mathbf k \parallel \mathbf n$. If there are
also unstable modes with more general wave vectors, they eventually
destroy the local Abelianization and lead to a saturation of
the exponential growth, which goes over into a slow linear one (Fig.\
\ref{fig3d}). As Fig.\ \ref{figcasc} shows,
at the level of the individual modes
this is accompanied by the build-up of an energy cascade where the energy
fed into the soft unstable modes by the
Weibel instability is distributed to
higher-momentum stable modes through 
non-Abelian self-interactions \cite{Arnold:2005ef,Arnold:2005qs}.

More recent lattice simulations using larger lattices and larger
(oblate) anisotropy Ref.~\cite{Bodeker:2007fw} 
found a continued exponential
growth of initially small perturbations
in the case of very strong momentum anisotropy, at least for
small initial field configurations, similar to the
behavior of the 1+1-dimensional case. With generic large
initial gauge field amplitudes, again a linear growth was observed.

\section{Quark-gluon plasma instabilities in Bjorken expansion}

In the above treatment of anisotropic plasmas, a space-time independent
background distribution was considered with a fixed momentum-space
anisotropy. A somewhat more realistic case appropriate for the
early stage of a quark-gluon plasma in formation after
a collision of large nuclei is given by a
distribution of hard massless particles which expands in one spatial
dimension (the beam axis) in a boost-invariant manner.

In the latter case it is convenient to switch to
comoving spacetime coordinates, proper time $\tau$ and rapidity $\eta$,
\be
x^\alpha=\left(\tau=\sqrt{t^2-z^2},x^i,\eta={\rm atanh}(z/t)\right)
\ee
with metric tensor $g_{\alpha\beta}=\rm diag(1,-1,-1,-\tau^2)$.
Momenta are parametrized by
\be
p^\alpha=|\mathbf p_\perp|
\left(\cosh(y-\eta),\cos\phi, \sin\phi,\tau^{-1}{\sinh(y-\eta)}\right)
\ee
with $y={\rm atanh}(p^0/p^z)$.
A boost-invariant, transversely isotropic, free-streaming background distribution is then given by
\be\label{faniso}
f_0(\1p,x)=f_{\rm iso}\left(\sqrt{p_\perp^2+p_\eta^2/\tau_{\rm iso}^2}\right)
=f_{\rm iso}\left(\sqrt{p_\perp^2+(p'^z\tau/\tau_{\rm iso})^2}\right)
\ee
where $p'_z$ is the boosted longitudinal momentum. Compared to (\ref{fxi})
we now have a proper-time dependent anisotropy parameter
\be
\xi(\tau)=(\tau/\tau_{\rm iso})^2-1,
\ee
such that there is an increasingly oblate momentum-space anisotropy 
for $\tau>\tau_{\rm iso}$.
This solves
\be
p^\mu\partial_\mu\, f_0(\1p,x)=p^\alpha\partial_\alpha f_0\Big|_{{\rm fixed}\; p^\mu}=0
\ee
because $(p^\alpha \6_\alpha) p_\eta(x)|_{{\rm fixed}\;p^\mu}=0$. (Note that
$p_\eta=-\tau^2 p^\eta$.)

Clearly, we need to start at some finite $\tau_0>0$ in order to avoid
a singularity at $\tau=0$. But a particle description of the initial
stage of a heavy-ion collision can anyway make sense only after
some finite time after the collision. In the so-called 
Color-Glass-Condensate (CGC) framework the earliest time when this is beginning to make sense is
provided by the inverse saturation scale $Q_s^{-1}$. In this framework, 
the initial hard-gluon density is given by \cite{Baier:2002bt}
\be
n(\tau_0)=c\, \frac{N_g Q_s^3}{4\pi^2 N_c\alpha_s (Q_s \tau_0)},
\ee 
where $c$ is 
the so-called gluon liberation factor. 
For definiteness, we shall 
assume $\tau_0 \sim Q_s^{-1}$
in what follows, and we shall also assume that $\tau_0 \gg \tau_{\rm iso}$,
i.e., that the momentum distribution is always strongly anisotropic
with positive $\xi$.

With increasing (proper) time, we have an increasing anisotropy with
more and more modes becoming unstable, but at the same time the
density decreases, which tends to diminish the growth rate.
The competition between these effects will of course modify the
evolution of non-Abelian plasma instabilities, but, at least
in the free-streaming idealization, the HL framework can be generalized
to study this case as well.



Since $p^\beta{\6_\beta} \,[\6_{(p)}^\alpha f_0(\mathbf p_\perp,p_\eta)]
|_{p^\mu=const.}=0$ (with index $\alpha$ upstairs!)
we can again solve the Vlasov equation
\be
p\cdot D\, \delta f_a(\1p,\1x,t)|_{p^\mu=const.}=
g p^\beta F_{\beta\alpha}^a \6_{(p)}^\alpha f_0(\1p,\1x,t),
\ee
by introducing auxiliary fields according to
\be
\delta f^a(x;p)=-g W^a_\alpha(\tau,x^{i},\eta;\phi,y)
\6_{(p)}^\alpha f_0(p_\perp,p_\eta)
\ee
with $W_\alpha^a$ satisfying
\be\label{vDWHEL}
v\cdot D\, W_\alpha(\tau,x^{i},\eta;\phi,y)|_{\phi,y}=v^\beta F_{\alpha\beta},
\ee
where we now define
\be
v^\alpha\equiv {p^\alpha\0 |{\mathbf p_\perp}|}
=\Bigl(\cosh(y-\eta),\cos\phi, \sin\phi,\frac{\sinh(y-\eta)}{\tau}\Bigr).
\ee

The induced current in the non-Abelian Maxwell equations, which in comoving coordinates now read 
\be
{1\0\tau}D_\alpha(\tau F^{\alpha\beta})=j^\beta, 
\ee
is given by
\be\label{jHEL}
\frac1\tau D_\alpha(\tau F^{\alpha\beta})=j^\beta(\tau,x^i,\eta)=
-{m^2\02}\int_0^{2\pi} {d\phi \0 2\pi}
\int_{-\infty}^\infty dy \,
v^\beta
\mathcal W(\tau,x^i,\eta;\phi,y)
\ee
with 
\be
\mathcal W=v^i W_i-{\tau\0\tau_{\rm iso}^2} \sinh(y-\eta)\,  W_\eta \,.
\ee


In the expanding case, already the effectively Abelian, weak-field
behavior is highly nontrivial, leading to complicated
integro-differential equations \cite{Romatschke:2006wg}, from
which one can extract that Weibel instabilities behave asymptotically
like
\be
\widetilde{A}^{i}(\tau,\nu)\sim \tau\, \,_2F_3
\left(\textstyle{\frac{3-s}{2}},\textstyle{\frac{3+s}{2}};2,2-i
\nu,2+i \nu;-\mu\tau
\right),\qquad s=\sqrt{1+4 \nu^2}
\label{Ai2F3}
\ee
where $\nu$ is the wave number corresponding to the rapidity variable $\eta$,
and 
\be
\mu={\pi\08} \tau_{\rm iso} m^2 .
\ee
For both $\nu$ and $\tau$ large, this leads to
\be
\widetilde{A}^{i}(\tau) \sim \tau^{1/4} \exp\left(2
\sqrt{
\mu\tau
}
\right),
\label{asymbeh}
\ee
which can be understood by the large-$\xi$ limit of $\gamma$ quoted at the
end of Sect.~\ref{secansio} together with the fact that the thermal
mass scale drops like the square root of hard particle density, which
behaves as $n\sim 1/\tau$. Such a behavior was found previously
in numerical CGC simulations that include small rapidity fluctuations
\cite{Romatschke:2006nk}.

\begin{figure}[tb]
\begin{center}
\begin{minipage}[t]{8 cm}
\centerline{\includegraphics[width=\textwidth]{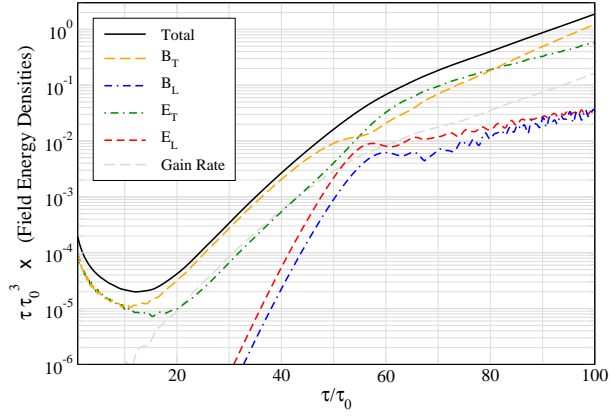}}
\end{minipage}
\begin{minipage}[t]{16.5 cm}
\caption{Results \cite{Rebhan:2008uj}
from a 1D+3V real-time lattice simulation of
non-Abelian plasma instabilities in Bjorken expansion, seeded by small initial
rapidity fluctuations with a spectrum modelled after Ref.~\cite{Fukushima:2006ax}. The plot shows
the proper-time dependence of
the total chromo-field energy density and its individual components
$\mathcal E=\mathcal E_{B_T}+\mathcal E_{E_T}+\mathcal E_{B_L}+\mathcal E_{E_L}
=\mathcal E_T+\mathcal E_L$
as well as the chromo-field energy gain rate $R$ defined by
$R={d{\cal E}}/{d\tau} + {2}{\cal E}_T/{\tau}$.
\label{fighel}
}
\end{minipage}
\end{center}
\end{figure}

In order to also investigate specific non-Abelian dynamics in the
HL framework,
it is again necessary to discretize space-time, now parametrized by
proper-time and rapidity coordinates, and the
non-compact velocity/momen\-tum-rapidity space parametrized
by $\phi$ and $y$. In fact, because 
the integrand in (\ref{jHEL}) is exponentially
suppressed at large $|y-\eta|$, only a finite rapidity interval
is required in numerical simulations.

Fig.~\ref{fighel} shows the results of a real-time lattice
simulation \cite{Rebhan:2008uj}
of the expanding HL equations (\ref{vDWHEL}) and (\ref{jHEL})
with small initial seed fields that are constant in transverse space
and supplied with an initial spectrum inspired by the
CGC scenario \cite{Fukushima:2006ax} and a gluon-liberation
factor \cite{Kovchegov:2000hz} $c=2\ln 2$ (which is also
close to the recent numerical results of Ref.~\cite{Lappi:2007ku}).
Given the findings in stationary anisotropic plasmas, the
growth of non-Abelian plasma instabilities in the
1+1-dimensional setting probably
gives an upper bound on more generic cases. Considering that
for $Q_s\simeq 1$ GeV for RHIC and $\simeq 3$ GeV for the LHC and that thus
the maximal time in Fig.~\ref{fighel} corresponds to some 20 fm/c for RHIC
and 7 fm/c for the LHC, one finds an uncomfortably long delay for the
onset of plasma instabilities at least for RHIC. Further studies
of more generic initial conditions (including then also strong
initial fields) are clearly called for to investigate this
situation in more generality. 

\section{Conclusion}

The extension of the HTL effective theory to anisotropic plasmas
leads to a (leading-order) theory of non-Abelian plasma instabilities,
which in the weak-field limit can be analysed by the study
of dispersion laws, but which in the strong-field case requires
numerical simulations on real-time lattices.
The latter have shown that contrary to initial expectations
non-Abelian plasma instabilities tend to saturate in the
nonlinear regime where fields are nonperturbatively large, but
not yet large enough for immediately modifying the anisotropic
background distribution of hard particles. However, a complete
isotropization scenario still needs to be worked out.
Moreover,
when boost-invariant expansion is taken into account,
there is an uncomfortably long delay for the onset of growth,
which seems too long for the environment provided by RHIC,
though not necessarily for the LHC, which may be where
wQGP physics eventually comes into its own.

It should be kept in mind that the above studies are based on
a truly weakly coupled situation with $g\ll 1$ and it is perhaps
not so surprising that at least RHIC physics lies outside of
the results obtained by extrapolation to $g\sim 1$, since
this is also the case when perturbative thermodynamic bulk results
are extrapolated to below 2--3 $T_c$.
Indeed, numerical simulations which do not separate collective HL physics
from other interactions have concluded that fast thermalization
through plasma instabilities may be indeed possible, see \cite{Dumitru:2006pz}.

Still, it remains a theoretical challenge to understand these
issues in the limit $g\ll1$ where a systematic perturbative analysis
should be possible, and, as said, perhaps this turns out to
find applications eventually at the higher scales to be probed by the LHC.

On the theoretical side, a systematic treatment of the HL effective
theory also offers new conceptual problems, as shown by its
application to heavy-ion observables such as energy loss
\cite{Romatschke:2003vc} 
and the
momentum broadening of jets
\cite{Romatschke:2006bb,Baier:2008js,Carrington:2008sp}.
There is clearly more work to be done, both numerically and analytically.

\section*{Acknowledgments}

I would like to thank the organizers of the 30th Course of the International School on Nuclear Physics in Erice for a wonderful meeting.
The work presented here was 
supported by FWF project no.\ P19526.


\end{document}